\documentclass[11pt]{article}
\usepackage{epsfig}
\usepackage{cite}
\usepackage{color}

\newcommand{\be}{\begin{equation}}
\newcommand{\ee}{\end{equation}}
\newcommand{\beeq}{\begin{eqnarray}}
\newcommand{\eeeq}{\end{eqnarray}}

\def\bea{\begin{eqnarray}}
\def\eea{\end{eqnarray}}

\def\xp{x_{{I\!\!P}}}

\def\qbar{\bar{q}}

\def\gev{\mbox{\rm GeV}}

\def\sigmahat{\hat{\sigma}}
\def\eto{{\rm e}}
\def\cbar{{\bar{c}}}
\def\bbar{{\bar{b}}}

\newlength{\dinwidth}
\newlength{\dinmargin}
\title{ \vspace {2.0cm}\bf
Heavy flavour production in DGLAP improved saturation model \vspace {0.5cm}}
\author{
K.~Golec-Biernat\footnote{e-mail: golec@ifj.edu.pl}\\
\textit{Institute of Nuclear Physics, Polish Academy of Sciences, 
        Cracow, Poland}\\
\textit{Institute of Physics, University of Rzesz\'ow, Rzesz\'ow, Poland}\\ \\
S.~Sapeta\footnote{e-mail: sapeta@th.if.uj.edu.pl}\\
\textit{M. Smoluchowski Institute of Physics, Jagellonian University, 
        Cracow, Poland}
}
\date{}

\begin{document}
\maketitle

\vspace*{0.5in}

\centerline{(\today)}

\vspace*{0.5in}

\begin{abstract}
\noindent
The charm and beauty quark production in deep inelastic scattering at low values of 
the Bjorken variable $x$  is considered in the DGLAP improved saturation model. 
After fitting parameters of the model to
the structure function $F_2$, the heavy quark contributions $F^{c\bar{c}}_2$ and $F^{b\bar{b}}_2$
are predicted. A good description of the data is found. 
Predictions for the longitudinal structure function $F_L$ and the diffractive structure function 
$F_2^D$ are also presented.

\end{abstract}

\newpage
\section{Introduction}

Among the color dipole models of deep inelastic scattering at small values of the Bjorken variable $x$, 
the saturation model 
\cite{Golec-Biernat:1998js,Golec-Biernat:1999qd} (GBW model)   turned out to be especially successful. 
It was able to describe both the  low $x$  structure function $F_2$ 
and the diffractive  structure function $F_2^D$ measured at HERA. 
From the theoretical side, the model has attracted a lot of attention since it grasps essential 
elements of parton saturation \cite{Gribov:1984tu, Mueller:1985wy,Mueller:1989st} incorporated in a relatively simple way. 
In particular, geometric scaling in DIS was predicted as a consequence of the assumptions 
on the saturation scale and scaling behaviour of the dipole cross section (which is a crucial element
of the dipole models) \cite{Stasto:2000er}. 
With the advent of more precise HERA data  \cite{Adloff:2000qk,Chekanov:2001qu,Breitweg:2000yn},
the GBW model needed improvement in order to provide better description of $F_2$  at 
large values of the photon virtuality
$(Q^2>20~{\rm GeV}^2)$. 
This was done by Bartels, Golec-Biernat and Kowalski (BGK)
by incorporating into the saturation model a proper gluon density evolving according to the DGLAP
evolution equations \cite{Bartels:2002cj}. This modification improves the small-$r$ part of the dipole 
cross section.

In the GBW saturation model the heavy quark contribution to $F_2$ was considered in the form of 
the $c\bar{c}$ pair production. 
This element, however, is missing in the BGK improvement.
The recent data from HERA  \cite{Adloff:2001zj,Aktas:2004az,Chekanov:2003rb} 
shows that the heavy quark contribution is of the size up to $30\%$ and  can by no means be neglected. 
Thus, the main goal of this analysis is to take into account heavy quark production 
in the DGLAP improved saturation model and confront it with the recent data. 
This analysis does not introduce new  parameters to those already present. 
Once the parameters of the dipole cross section are determined from a fit to the total structure function $F_2$, 
they can be  used to \emph{predict} the charm and beauty contributions  
$F^{c\bar{c}}_2$ and $F^{b\bar{b}}_2$, which are not fitted separately. 
In addition,  the longitudinal structure function $F_L$ 
and the diffractive structure function $F^{D}_2$ can  also be predicted.

It should be stressed that other dipole models are also successful in the description of the  DIS data.
These include the model of Forshaw and Shaw \cite{Forshaw:1999uf,Forshaw:1999ny,Forshaw:2004vv} based on the Regge-like ideas combined with the idea of saturation, 
the model of Iancu, Itakura and Munier \cite{Iancu:2003ge} closely related to the theory of the color
glass condensate \cite{McLerran:1993ni,McLerran:1993ka,Jalilian-Marian:1997gr,Jalilian-Marian:1997dw,Weigert:2000gi,Iancu:2000hn,Ferreiro:2001qy,Iancu:2001ad}
and the model of McDermott, Frankfurt,  Guzey and Strikman \cite{McDermott:1999fa} which is 
close in spirit to our analysis. 
Quite recently, a new analysis of Kowalski, Motyka and Watt \cite{Kowalski:2006hc} appeared
which addresses a very important problem of the impact parameter dependence of the dipole scattering amplitude.
For more details on these models see recent review \cite{Forshaw:2006np}.

The outline of this paper is as follows. In Section~\ref{sec:model} we recall the main features of 
the BGK saturation model. In Section~\ref{sec:fit} we describe fits with heavy flavours to 
the data on $F_2$ from HERA.  In Section~\ref{sec:critical} we discuss main features of parton saturation, namely
the critical line and saturation scale.
Based on the results of the fits, in  Section~\ref{sec:heavy} we predict
the charm and beauty contributions to $F_2$ and 
the longitudinal structure function. In Section~\ref{sec:diff} we  discuss 
the diffractive structure function.
Finally, conclusions are presented in Section~\ref{sec:conclusions}.

\section{The BGK saturation model}
\label{sec:model}

In the dipole picture of DIS at small values of $x$  \cite{Nikolaev:1990ja,Nikolaev:1991et,Mueller:1993rr,Mueller:1994jq,Mueller:1994gb}, 
the virtual photon--proton scattering
cross sections $\sigma^{\gamma^*p}_{T,L}$ from transverse (T) and longitudinal (L) photons are given as  a convolution of  the photon wave functions $\psi_{T,L}$ 
and the dipole cross section $\sigmahat$,
\be\label{eq:sigmatl}
\sigma^{\gamma^*p}_{T,L}(x,Q^2)\,=\,\sum_{f=1}^{N_f}\int d^2r\, dz\,
|\psi_{T,L}(r,z,Q^2;m_f,e_f)|^2\,\,\sigmahat(x,r)\,,
\ee
where $f$ denotes flavour of the quark--antiquark pairs (dipoles)
interacting with the proton ($m_f$ and $e_f$ are quark mass and electric charge, respectively). 
The integration is performed over the $q\qbar$ transverse separation vector $r$  
and the light-cone momentum fraction of
the virtual photon  $z$ carried by the quark or antiquark.
The proton structure function $F_2$ is proportional to the sum of the two cross sections
\be
F_2=\frac{Q^2}{4\pi^2\alpha_{em}}\,(\sigma^{\gamma^*p}_{T}+\sigma^{\gamma^*p}_{L})\,.
\ee

In the BGK model \cite{Bartels:2002cj} the dipole cross section was assumed  in the 
form 
\be
\label{eq:sighatnew}
\hat\sigma (x,r)\,=\,\sigma_0\,\left\{
1\,-\,\exp\left(-\frac{\pi^2\,r^2\,\alpha_s(\mu^2)\,xg(x,\mu^2)}
{3\,\sigma_0}\right) \right\}\,,
\ee
which recovers the well known result of perturbative QCD \cite{Frankfurt:1996ri} that 
for small dipole sizes the dipole cross section is proportional
to the gluon distribution $g(x,\mu^2)$ with the scale $\mu^2\sim 1/r^2$. 
Indeed, expanding (\ref{eq:sighatnew}) in powers of $r^2$, we find to the first order  
\be
\label{eq:smallr}
\hat\sigma(x,r) \,\simeq\,
\frac{\pi^2}{3}\,r^2\,\alpha_s\,xg(x,\mu^2)\,.
\ee
The feature that $\hat\sigma(x,r)$ vanishes in the limit
$r\to 0$ is called colour transparency and results from gauge invariance of QCD.
The gluon distribution evolves with the scale according to the leading order DGLAP evolution equations. 
Since the model is applied in the low~$x$ region, quarks are neglected in the evolution.
Inspired by the MRST parametrisation \cite{Martin:2002dr},
the initial gluon distribution for the DGLAP evolution is taken in the form
with two parameters, $A_g$ and ${\lambda_g}$, to be  determined from fits to data
\be
\label{eq:gluon}
xg(x,Q_0^2)\,=\,A_g\,x^{\lambda_g}\,(1-x)^{5.6}\,.
\ee
The  initial scale $Q^2_0$  is conventionally set to $1~\gev^2$

The  behaviour of the dipole cross section (\ref{eq:sighatnew})  at large $r$
has not been changed with respect 
to the GBW model in which the dipole cross section saturates at the maximal value $\sigma_0$
(also to be determined from fits to data). 
This key assumption is in agreement with unitarity of the cross section $\sigma^{\gamma^*p}$.
The large dipole sizes, however,  are the cause of troubles for the scale
of the gluon distribution in Eq.~(\ref{eq:sighatnew}) since $\mu$  may become smaller
than $\Lambda_{\rm QCD}$. In order to avoid this we freeze the scale at some minimal
value $\mu_0 \gg\Lambda_{\rm QCD}$. Thus we assume that 
\be
\label{eq:scale}
\mu^{2}\,=\,\frac{C}{r^2}\,+\,\mu_0^2\,,
\ee
which introduces two additional parameters, $C$ and $\mu_0^2$, to be found from fits to data. 
For $r\to 0$ the presence
of $\mu_0^2$ becomes irrelevant and we recover the relation $\mu^2\sim 1/r^2$.

On the whole, the BGK model has five parameters to be fitted: 
the dipole cross section bound $\sigma_0$ and  
the four parameters of the gluon distribution: $A_g,\,\lambda_g,\,C$ and $\mu_0^2$. 
In comparison to the GBW saturation model, there
are two new parameters introduced by the scale (\ref{eq:scale}).

\section{Fit description}
\label{sec:fit}

We performed fits with  the charm and beauty contributions in the sum in Eq.~(\ref{eq:sigmatl})
using recent data on the proton structure function $F_2$ 
from H1 \cite{Adloff:2000qk} and ZEUS \cite{Chekanov:2001qu, Breitweg:2000yn}.  We considered
only the data points with $x \leq 0.01$ and $Q^2 \ge 0.04\, {\rm GeV}^2$.
The number of experimental points in such a case was equal to $288$.  
The statistical and systematic errors were added in quadrature. 
Moreover, the H1 data were multiplied by a factor $1.05$ to account for slightly 
different normalisation of the H1 and ZEUS data sets. 
Similarly to the analyses \cite{Golec-Biernat:1998js,Bartels:2002cj}, we also modified the argument in the dipole cross section
$\hat{\sigma}(x,r)$ in the heavy flavour contributions,
\be
\label{eq:xbmod}
x~\to~x\left(1+\frac{4\/m_f^2}{Q^2}\right)\,=\,\frac{Q^2+4\/m_f^2}{Q^2+W^2},
\ee
where $W$ is invariant energy of the $\gamma^*p$ system.
This modification  takes into account that the  heavy quark masses play the role
of the hard scale when $Q^2\ll m^2_{c,b}$.

We took typical values of the heavy quark
masses, $m_{\rm c}=1.3~{\rm GeV}$ and  $m_{\rm b}=5.0~{\rm GeV}$, but 
the light quarks stayed massless. Therefore, we excluded the photoproduction
point $Q^2=0$ from considerations 
since in the  dipole models $\sigma^{\gamma p}$ depends logarithmically on 
the quark mass in the  limit $Q^2\to 0$.
A comment is in order at this point. We performed preliminary fits with the light quark mass $m_q =  140\ {\rm MeV}$ taken from \cite{Golec-Biernat:1998js}. However, we did not find the value of the photoproduction cross section measured at HERA ($174\ {\rm \mu b}$)\cite{Chekanov:2001gw}. Thus the description of photoproduction would require fine-tuning of the light quark mass in the region where perturbative QCD, which is the basis of the dipole models, is not valid.
Nevertheless, in the case of the heavy flavour production, when the quark mass is
the hard scale, the predictions for $Q^2=0$ could be done.

Before we present the results of our fits, let us mention that 
the GBW model with heavy flavours
fitted to the new data \cite{Adloff:2000qk,Chekanov:2001qu, Breitweg:2000yn} 
gives $\chi^2/{\rm ndf} = 2.32$.
As it has already been pointed out in \cite{Bartels:2002cj},
this rather poor fit quality results from the  lack of the proper DGLAP evolution
of the gluon distribution in the GBW model.

We performed two fits with the dipole cross section (\ref{eq:sighatnew}), taking 
into account
the charm and beauty contribution in addition to the three light quarks. In the first fit only  charm
was considered while in the second one both heavy flavours were present.
We set the number of active flavours in $\alpha_s$ to $4$ and $5$, respectively and the value of 
$\Lambda_{\rm QCD} = 300\ {\rm MeV}$ in both cases. 
The fit results for the five parameters of the model, $\sigma_0$, $A$, $\lambda_g$, $C$ and $\mu^2_0$, 
are presented in Table~\ref{fit_results}. 
In the last row we recall from \cite{Bartels:2002cj} the parameters of the BGK fit 
with the light quarks only.

As we see, the value of $\chi^2/{\rm ndf}$ is still good for the fits with heavy flavours.
The gluon parameters differ significantly from the light quark fit. 
In particular, the power $\lambda_g$ is negative
which means that the initial gluon distribution (\ref{eq:gluon}) grows with decreasing $x$, 
in contrast to the fit with light quarks only when the gluon distribution is valence--like
($\lambda_g$ is positive). 
With the found  gluon density, the  total proton momentum fraction carried by gluons is around
$25\%$ at the inital scale $Q^2_0=1~{\rm GeV}^2$.

\begin{table}[t]
\begin{center}
\begin{tabular}{|c||c|c|c|c|c||c|} \hline 
& $\sigma_0\,$[mb] & $A_g$ & $\lambda_g$ & $C$ & $\mu^2_0$ & $\chi^2/ndf$
\\ \hline  \hline 
light + c + b                  & 22.7 &~1.23~&~- 0.080~~&~0.35~&~1.60~&~1.16~
\\ \hline
light + c                      & 22.4 &~1.35~&~- 0.079~~&~0.38~&~1.73~&~1.06~
\\ \hline\hline
light                          & 23.8 &13.71 &~~0.41~&~11.10~&~0.52~& 0.97
\\ \hline
\end{tabular}
\end{center}
\caption{The parameters of our fits with heavy quarks to the $F_2$ data. 
The results of the BGK fit from\cite{Bartels:2002cj} 
 with massless light quarks only are given for the reference in the last row.}
\label{fit_results}
\end{table}

\begin{figure}[p]
\begin{center}
\rotatebox{270}{\psfig{figure=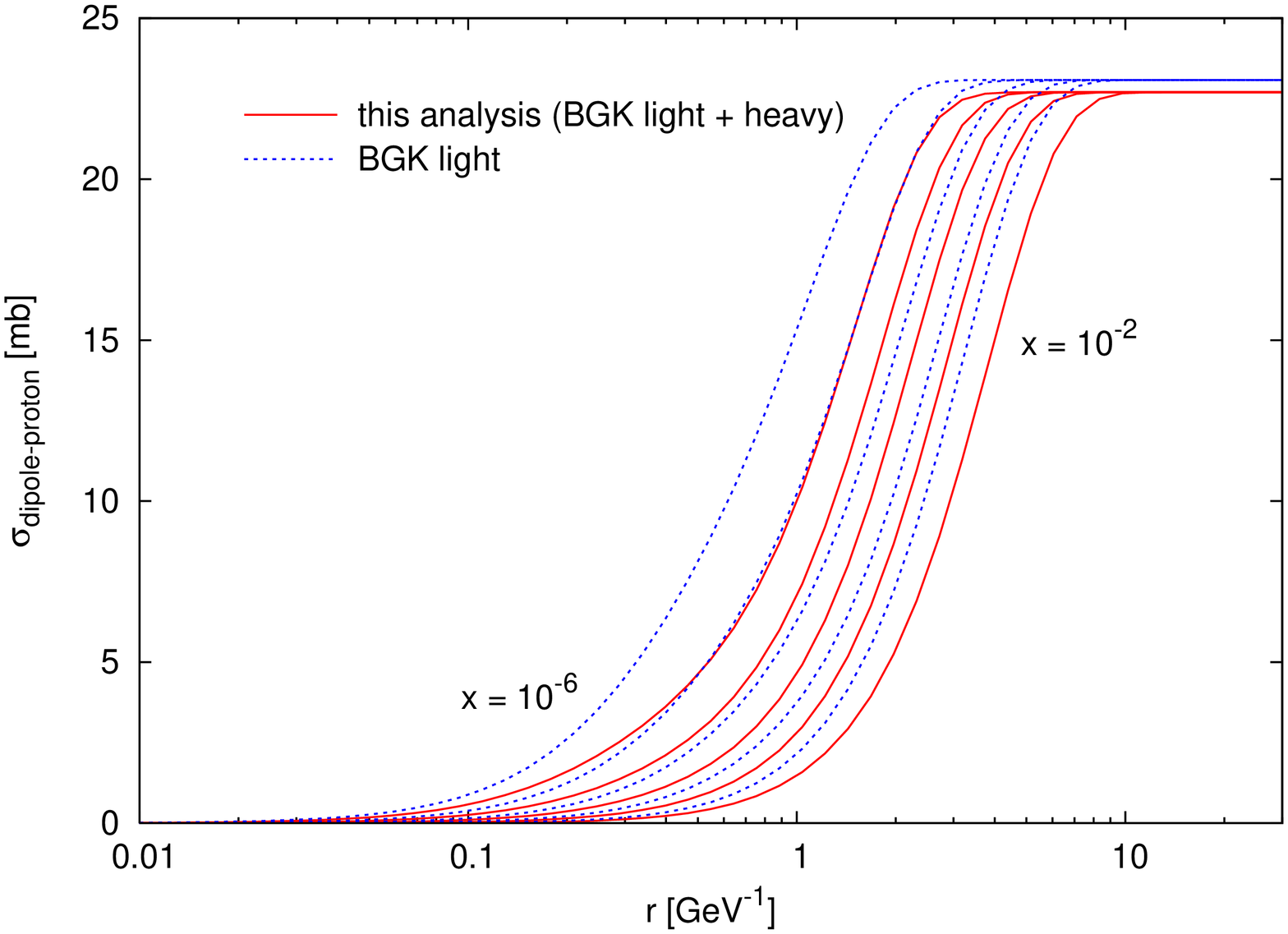,height=5.2in}}
\caption{The dipole cross section in the BGK model with and without heavy quarks (solid and dashed lines, respectively) for $x= 10^{-2} \dots 10^{-6}$.}
\label{fig:dipole_cs}
\end{center}
\end{figure}

In Fig.~\ref{fig:dipole_cs} 
we show the comparison of the dipole cross sections from the present analysis with heavy quarks (solid lines) and the BGK analysis \cite{Bartels:2002cj} with light quarks only (dashed lines). The effect of heavy quarks is seen in the shift of the dipole cross section towards larger values of $r$, which means that for a given dipole size saturation occurs at lower $x$ (higher energy). 
This effect was also observed in the GBW analysis \cite{Golec-Biernat:1998js}.

It should also be mentioned that the presence of heavy quarks in the DGLAP improved model 
cures the pathological behaviour of the dipole cross section found in \cite{Bartels:2002cj}
in  the fit with massless quarks (FIT 2). 
This is why we choose the BGK fit with massive light quarks (FIT 1) for the comparison in  Fig.~\ref{fig:dipole_cs}.

\section{Critical line and saturation scale}
\label{sec:critical}

\begin{figure}[p]
\begin{center}
\psfig{figure=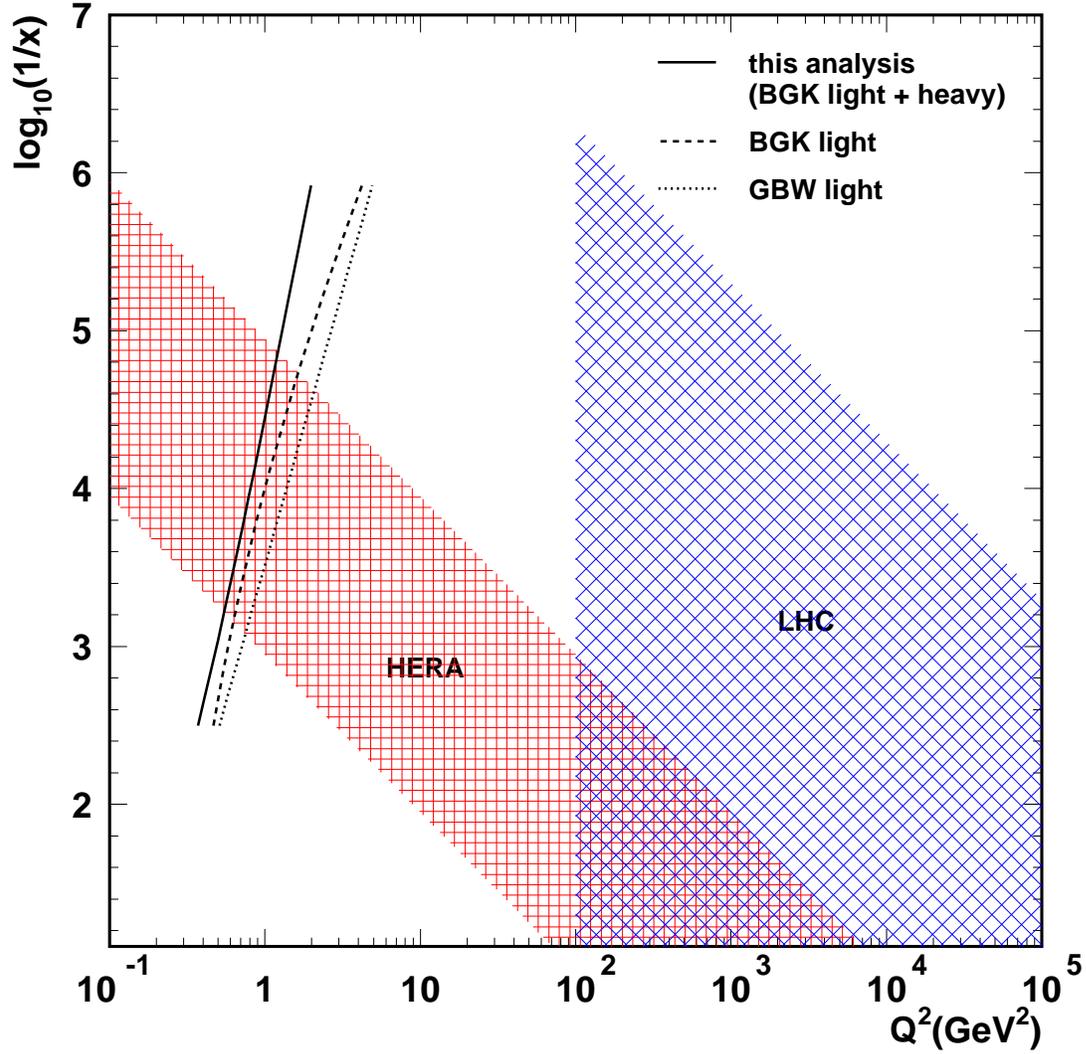,height=6.1in}
\caption{The critical line in the $(x, Q^2)$-plane from various saturation models indicating the position of the saturation region (to the left of these lines). The shaded areas show the acceptance regions of HERA and the LHC. The latter region corresponds to the production of an object with the minimal mass squared $Q^2= 100\ {\rm GeV^2}$ \cite{Martin:1999ww}.}

\label{fig:crit_line}
\end{center}
\end{figure}

The shift of the dipole cross section towards larger values of $r$ has direct impact on the position of the  critical line
which marks the transition to the saturation region in the $(x,Q^2)$--plane.
The line is defined by the condition
\begin{equation}
\hat{\sigma}_0(x,\bar{r}) = a\, \sigma_0,
\end{equation}
with the constant $a$ of the order of unity and 
the dipole size taken at its characteristic value $\bar{r}=2/Q$. 
In the GBW and BGK analyses the critical line was defined by 
the condition 
that the argument of the exponent in the dipole cross section equals 1,
which  corresponds to $a=1-\eto^{-1}\approx 0.63$. 
With the form (\ref{eq:sighatnew}), the following implicit relation
between  $x$ and $Q^2$ is found in such a case
\begin{equation}
\frac{4\pi^2}{3\sigma_0 Q^2}\, \alpha_s({\mu^2})\, xg(x,\mu^2) = 1\,,
\end{equation}
where the scale $\mu^2=CQ^2/4+\mu^2_0$. 
This equation can be solved numerically to obtain
the  critical line  shown in Fig.~\ref{fig:crit_line} as the solid line. 
The saturation effects are important to the left of this line. For the comparison, we also show
the critical lines from the BGK and GBW analysis
with light quarks only.
We observe that the presence of heavy quarks shifts the critical line towards smaller values of $Q^2$. 
This means that for a given  $Q^2$ we need lower $x$ in order to stay in the domain 
where the saturation effects are important. In other words, heavy quarks make saturation
more difficult to observe at present  and also future colliders,  which is indicated in Fig.~\ref{fig:crit_line} by the acceptance regions of HERA and the LHC.

An important element of the description of 
parton saturation in the GBW model \cite{Golec-Biernat:1998js} was the saturation scale,
$Q_s(x) = Q_{s0}\, x^{-\lambda}$, built in the dipole cross section
\be
\label{eq:sighatnew1}
\hat\sigma(x,r)\,=\,\sigma_0\,\left\{
1\,-\,\exp\left(-{\textstyle\frac{1}{4}} r^2 Q^2_s(x)\right) \right\}.
\ee
This form of the dipole cross section features the scaling behaviour , $\hat\sigma(x,r)=\hat\sigma(rQ_s(x))$, 
which  for massless quarks leads to geometric scaling of the total cross section,
\be
\sigma^{\gamma^*p}=\sigma^{\gamma^*p}(Q^2/Q^2_s(x)),
\ee
found to a good accuracy in the small $x$ HERA data \cite{Stasto:2000er,Marquet:2006jb}. Notice that the critical line
in the GBW model is defined as a simple condition: $Q^2=Q^2_s(x)$.

\begin{figure}[p]
\begin{center}
\rotatebox{270}{\psfig{figure=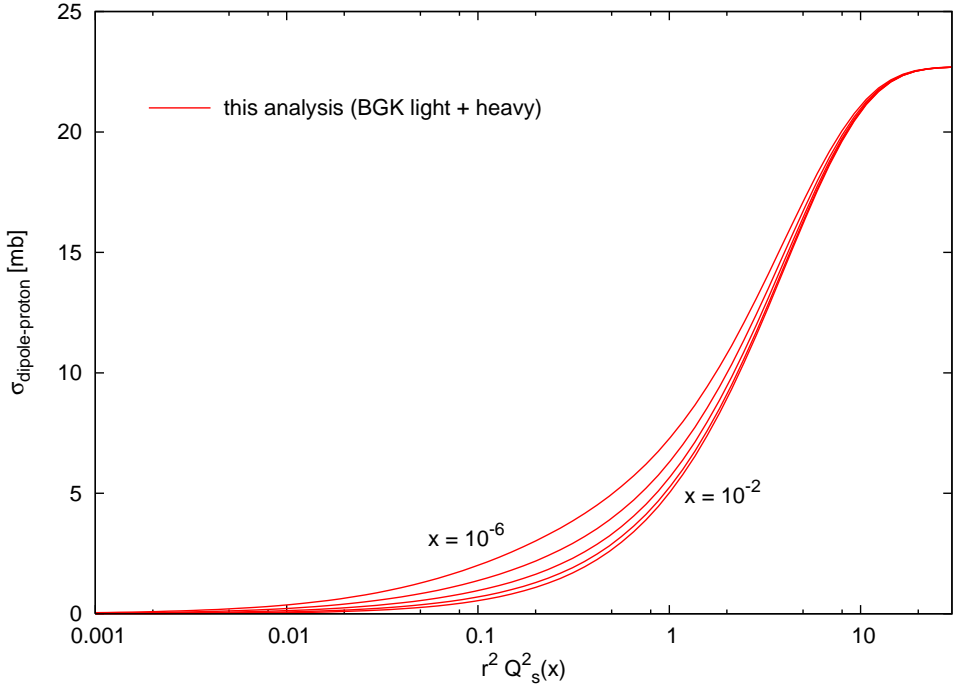,height=5.2in}}
\caption{The dipole cross section in the BGK model with heavy quarks as a function of the scaling variable $r^2 Q^2_s(x)$ with the saturation scale given by Eq.~(\ref{eq:BGKsc}).
Geometric scaling is preserved for moderate dipole sizes, while it
is broken for small values of $r$ because of the DGLAP modification.}
\label{fig:dipol_scaling}
\end{center}
\end{figure}

The BGK modification
of the saturation model seems to  abandon these important elements. Fortunately, this is not the case
since in the fits with heavy flavours 
the value of $C$ in the gluon scale (\ref{eq:scale}) is small whereas 
$\mu^2_0\approx 1.6~\gev^2$ is relatively large. It means that for not too small $r$, the dipole cross sections 
(\ref{eq:sighatnew}) effectively features the scaling at large values of $r$ 
with the  saturation scale proportional 
to the gluon distribution  at the scale $\mu_0^2$
\begin{equation}
\label{eq:BGKsc}
Q^2_s(x) \simeq \frac{4\,\pi^2}{3\,\sigma_0}\, \alpha_s({\mu^2_0})\, xg(x,\mu^2_0)\,.
\end{equation}
Indeed, as we see in Fig.~\ref{fig:dipol_scaling}, which shows the dipole cross section 
as a function of the scaling variable $r^2 Q^2_s(x)$, 
geometric scaling is preserved for moderate values of $r$. It is broken, however,  for small dipole sizes due to 
the DGLAP evolution of the gluon in the dipole cross section. 

It is important to notice that from the theoretical perspective, based on the theory of the color glass
condensate (see \cite{Iancu:2006qi} for most recent review), our considerations are confined to the {mean field approximation} which neglects
the effects of fluctuations in the number of color dipoles in the proton wave function. These effects
are important in the low density region of dipoles with very small sizes. For sufficiently small $x$
(large rapidity $Y=\ln(1/x)$), the fluctuations wash out geometric scaling leading to {\it diffusive scaling} \cite{Mueller:2004se,Iancu:2004es,Munier:2005re,Soyez:2005ha,Enberg:2005cb,Hatta:2006hs,Brunet:2005bz,Marquet:2006xm}.
The DIS cross sections in this case become functions of the variable $\ln[Q^2/Q_s^2(Y)]/\sqrt{DY}$ rather than
$Q^2/Q_s^2(Y)$. The  discussion of the new type of scaling is, however,  beyond the scope of the presented analysis.

\section{Predictions for inclusive structure functions}
\label{sec:heavy}

Once the parameters of the dipole cross sections (\ref{eq:sighatnew}) are determined from the fit to the $F_2$ data,
various inclusive structure functions can be predicted. 
In particular, the heavy quark contributions, $F_2^{c\bar{c}}$ and $F_2^{b\bar{b}}$, 
can be found from the flavour decomposition in  Eq.~(\ref{eq:sigmatl})
\be
F_2=F_2^{light}+F_2^{c\bar{c}}+F_2^{b\bar{b}}\,.
\ee
The dependence on flavour comes through electric charge  $e_f$ 
and mass $m_f$ in the photon wave functions $\psi_{T,L}$. For heavy quarks the mass  also enters the dipole cross section 
through the modified value of the Bjorken variable (\ref{eq:xbmod}).
This allows to compute the $c\cbar$ and $b\bbar$ photoproduction cross sections.
For the HERA energy $W=209~\gev$,  we found $19.3~\mu b$ and $0.7~\mu b$, respectively. To our astonishment,
substituting the mass $m_q=140\ {\rm MeV}$ for the three light quarks to the formula for $\sigma^{\gamma p}$
and performing then the photoproduction limit $Q^2\to 0$, we found $177\ {\rm \mu b} $  which agrees with
the measured value $174\ {\rm \mu b}$ up to the experimental errors.

\begin{figure}[p]
\begin{center}
\psfig{figure=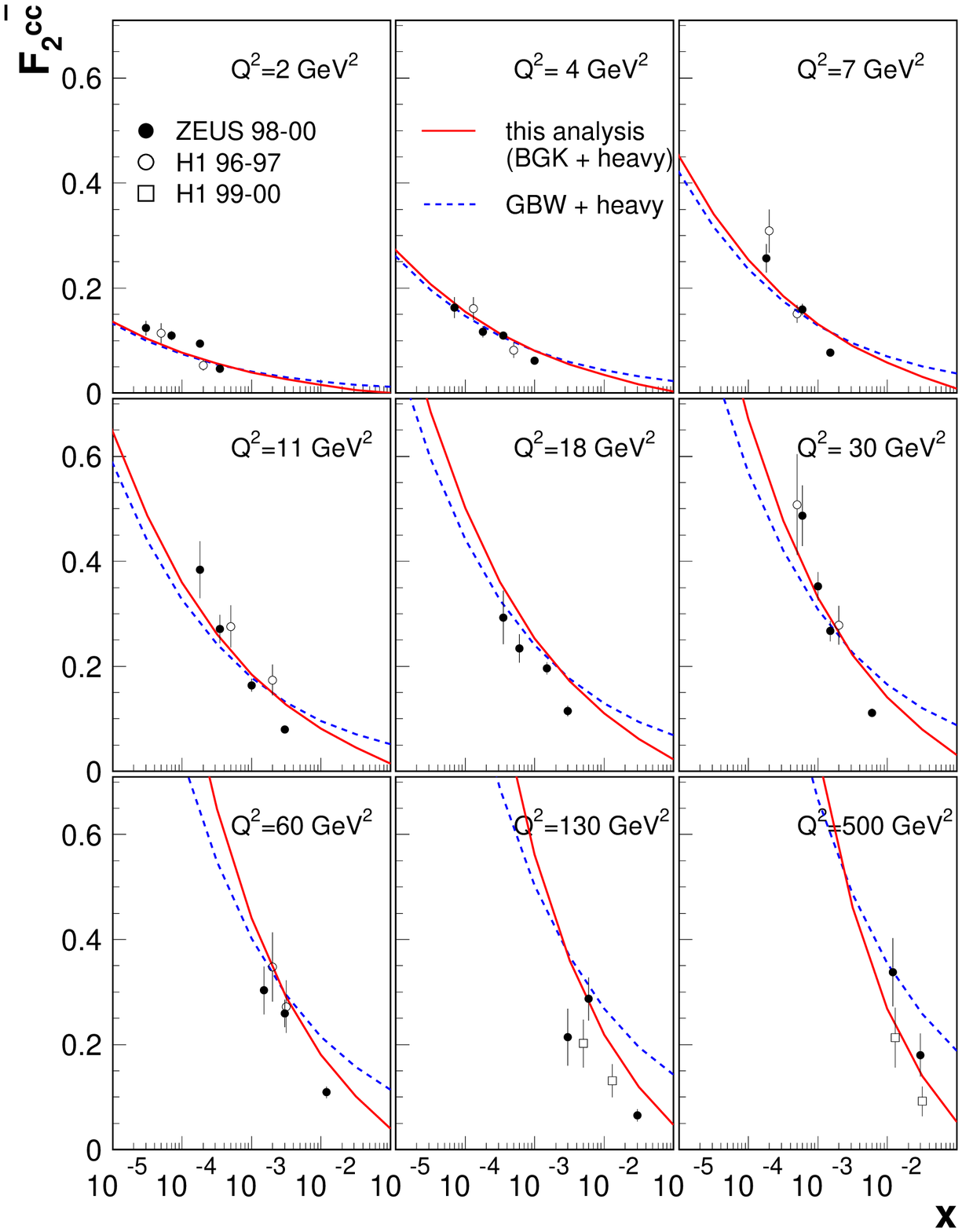,height=6.1in}
\caption{Predictions for the charm structure function $F^{c\bar{c}}_2$ in the BGK   model 
with heavy quarks (solid lines). The predictions in the GBW model \cite{Golec-Biernat:1998js}
are shown for reference (dashed lines).} 
\label{fig:f2charm}
\end{center}
\end{figure}

\begin{figure}[p]
\begin{center}
\psfig{figure=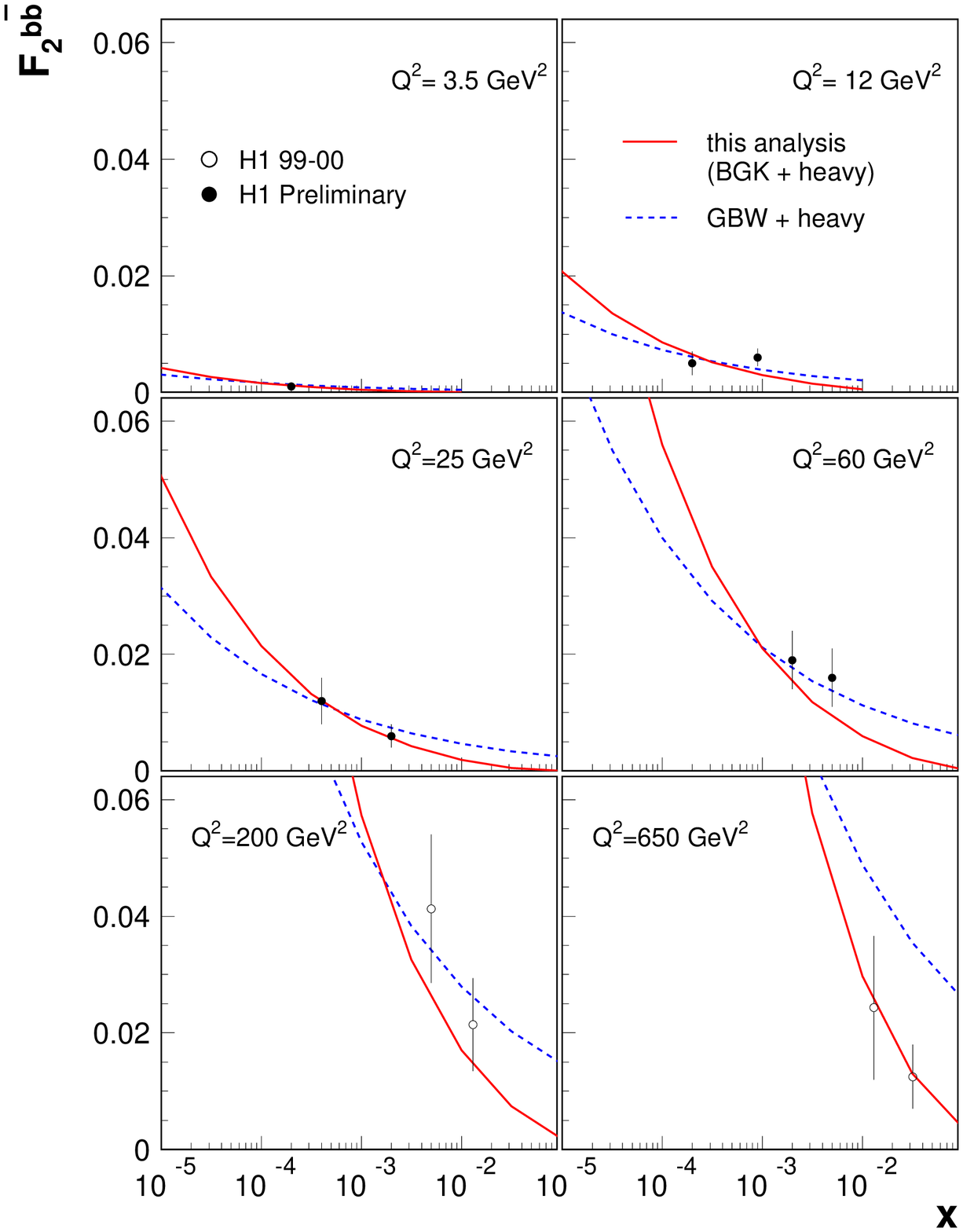,height=6.1in}
\caption{Predictions for the beauty structure function $F^{b\bar{b}}_2$ in the BGK   model 
with heavy quarks (solid lines). The predictions in the GBW model \cite{Golec-Biernat:1998js}
are shown for reference (dashed lines).} 
\label{fig:f2beauty}
\end{center}
\end{figure}

The predictions for $F_2^{c\bar{c}}$ and $F_2^{b\bar{b}}$  computed with the parameters from
the first  line of Table~1 are presented as the solid lines in Figs.~\ref{fig:f2charm} and \ref{fig:f2beauty},
respectively. For the comparison, we present the predictions of
the GBW model without the DGLAP modification (dashed lines). 
We see very good agreement with the data from HERA, both in the normalisation
and the slope in $x$, in contrast to the GBW results
which overshoot the data at large values of $Q^2$. 
Thus, the presence of the DGLAP evolution in the BGK model 
is crucial for the correct predictions at large $Q^2$. 
This can be understood by analysing the contribution of different dipole sizes 
to the heavy quark structure functions. For $Q^2\gg 4\/m_f^2$,  
they are mostly sensitive to the small--$r$ part of the dipole cross section
($r \approx 2/Q$) which is strongly modified by the DGLAP improvement.

\begin{figure}[p]
\begin{center}
\rotatebox{270}{\psfig{figure=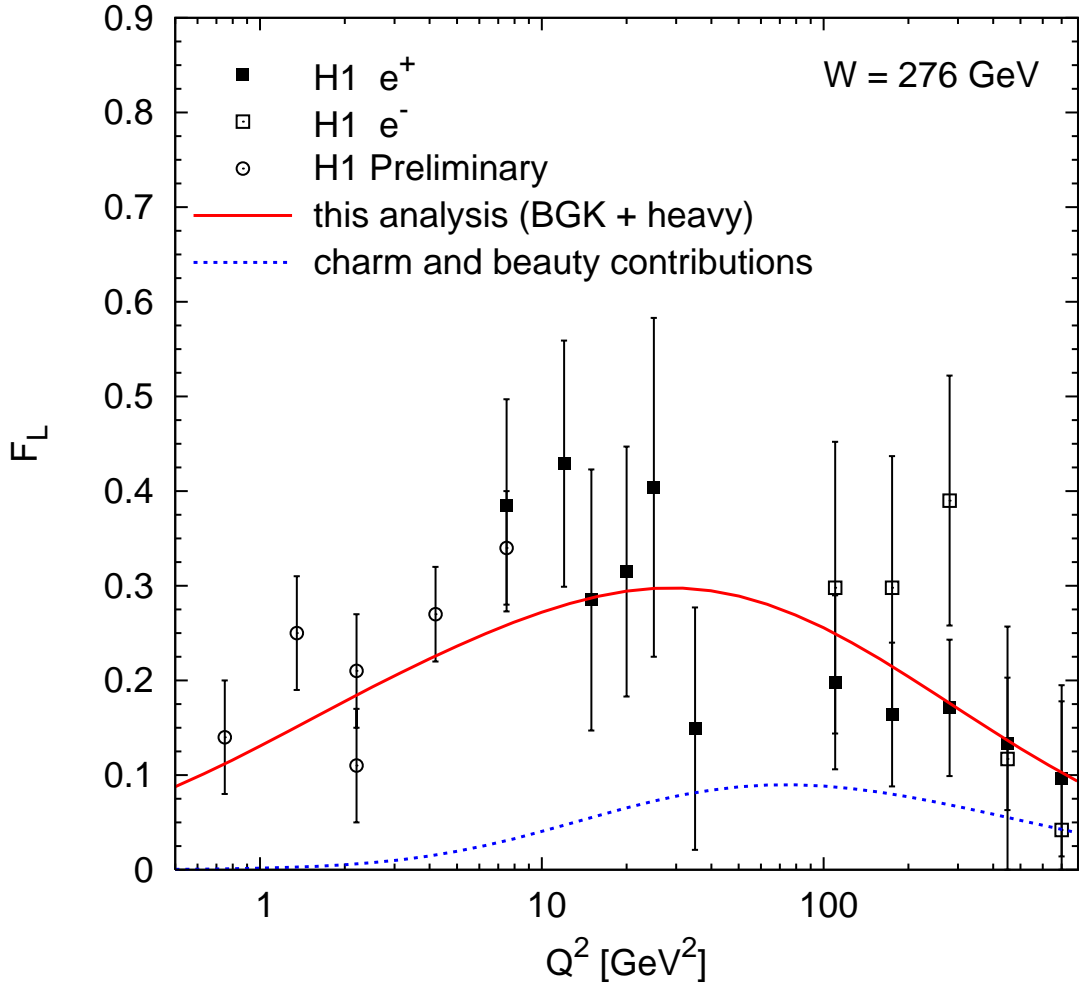,height=6.0in}}
\caption{The longitudinal structure function predicted in the BGK model with heavy quarks together with the 
H1 estimations  for various $Q^2$ at constant energy W = 276~\gev.}
\label{fig:flq2}
\end{center}
\end{figure}

In Figure~\ref{fig:flq2} we present the longitudinal structure function 
from our analysis (solid line), plotted against $Q^2$ 
for $W=276~\gev$. 
The experimental points represent the H1  estimations of $F_L$  
\cite{Adloff:2000qk, Adloff:2003uh, Lobodzinska:2004ig}. 
Reasonable agreement is found, however, the estimation errors are
too large to draw firm conclusions.
We also show the charm and beauty contribution to $F_L$ (dashed line). 
We observe that in our analysis
heavy quarks are important for large values of $Q^2$ 
while for $Q^2<10\ {\rm GeV}^2$ they may safely be neglected.

\section{DIS diffraction}
\label{sec:diff}

Deep inelastic diffraction is an important test of the dipole models \cite{Golec-Biernat:1999qd,Forshaw:2004xd,Munier:2003zb}. The dipole cross
section from the inclusive analysis can be applied to the description of the diffractive
structure function. Following the approach presented in \cite{Golec-Biernat:1999qd},
the diffractive structure function is the sum of three contributions
\be
F_2^{D(3)}(\xp,\beta,Q^2)\,=\,F_T^{q\qbar}+
F_L^{q\qbar}+F_T^{q\qbar g}\,,
\ee
which correspond to the diffractive systems  composed of the $q\qbar$ and $q\qbar g$ dipoles with invariant mass $M$,
produced by the transversely and longitudinally polarised virtual photons. 
Their interaction with the proton is described by the dipole cross section $\sigmahat(\xp,r)$. 
Note that
\be
\xp=\frac{Q^2+M^2}{Q^2+W^2}=\frac{x}{\beta}
\ee
is substituted instead of the Bjorken $x$ as an argument of $\sigmahat$.
The longitudinal component $F_L^{q\qbar g}$ is suppressed by $1/Q^2$ and is negligible
in the region of small values of $\beta=Q^2/(Q^2+M^2)$, \emph{i.e.} for large diffractive masses $M^2\gg Q^2$,
where the transverse $q\qbar g$ component dominates.

The $q\qbar g$  diffractive amplitude was computed in \cite{Golec-Biernat:1999qd} in the two-gluon exchange approximation with strong ordering of transverse momenta of  the $q\qbar$ pair and the gluon.
Each contribution is proportional to the inverse of the diffractive slope $B_D$ for which we took the experimental value $B_D=6~\gev^{-2}$. In addition, the $q\qbar g$ component is proportional to the strong coupling  
which we took a typical $\alpha_s = 0.2$. However, this is not quite clear which value of $\alpha_s$ should be used, thus the  normalisation of this term is rather uncertain.

In Fig.~\ref{fig:f2diff} we show the comparison of the diffractive structure function $F_2^{D(3)}$ computed in
our analysis with the data from H1 \cite{unknown:2006hy,unknown:2006hx} and  ZEUS \cite{Chekanov:2004hy, Chekanov:2005vv}.
However,  these data  are obtained for different upper limits on the mass of the proton dissociation system,  $M_Y<2.3$~GeV for ZEUS and $M_Y<1.6$~GeV for H1. To account for this, the ZEUS data in Fig.~\ref{fig:f2diff} is
multiplied by the global factor $0.86$ in accordance with \cite{unknown:2006hy}. 
At the same time we multiplied our predictions for the nondissociated proton by $1.23$.
This number, taken from \cite{unknown:2006hy}, is the correction factor which takes into account the proton dissociation up to the mass $M_Y<1.6$~GeV. Thus, both the data and the theoretical predictions are normalised to the same experimental situation.
The parameters from the first line of Table~1 were used in the dipole cross section for the computation of $F_2^{D(3)}$. As we see from Fig.~\ref{fig:f2diff}, reasonable agreement with the data is found. 

In contrast to the inclusive case, the $c\bar{c}$ and $b\bar{b}$ contributions to the diffractive structure function
are negligible due to the phase space restrictions for the diffractive system. The analysis of the   
diffractive $c\cbar g$ production, however,  
deserves more detailed considerations based on the seminal work \cite{Wusthoff:1999cr,Bartels:2002ri}.

\section{Conclusions}
\label{sec:conclusions}

We discussed the production of the charm and beauty flavours in the DGLAP improved saturation model \cite{Bartels:2002cj}. We fitted the parameters of the model to the recent $F_2$ data from HERA,
taking into account the heavy quark contributions in the theoretical formula for $F_2$.
We found a good quality fit with $\chi^2/{\rm ndf}$ 
close to unity. 
Thus we conclude that the successful description of the inclusive $F_2$ data at low $x$ which was found for the BGK model with light quarks  is preserved when heavy flavours are considered. There are, however, some differences introduced by heavy quarks. First of all, the parameters vary significantly for the models with and without heavy flavours. This results in the shift of the dipole cross section towards larger values of the dipole sizes $r$ with respect to the 
light quark case. As a consequence, the critical line in $(x,\ Q^2)$--plane moves in the direction of smaller 
values of $Q^2$ which makes saturation more difficult to observe.

The new predictions provided by our analysis concern the charm and beauty structure functions $F^{c\bar{c}}_2$ and $F^{b\bar{b}}_2$. We found very good agreement with H1 and ZEUS data in all $Q^2$ bins. The significant improvement of the slope in $x$ for high $Q^2$ with respect the GBW model is attributed to the DGLAP evolution.

Other structure functions were also determined. For the longitudinal structure function $F_L$, we found agreement with the H1 estimations. However, large estimation errors prevent from making more precise statement. The comparison with direct measurements, which are planned by H1 and ZEUS collaborations, will be particularly interesting.
For the diffractive structure function $F^{D(3)}_2$, we observe reasonable agreement with the H1 and ZEUS data after a proper renormalisation of the data points.

Finally, we discussed the issues related to the  essential features of parton saturation like
the saturation scale and geometric scaling. We showed that saturation scale is effectively present
in the DGLAP improved model being related to  the gluon distribution at the frozen value of the
factorisation scale $\mu_0^2$.

\begin{figure}[p]
\begin{center}
\psfig{figure=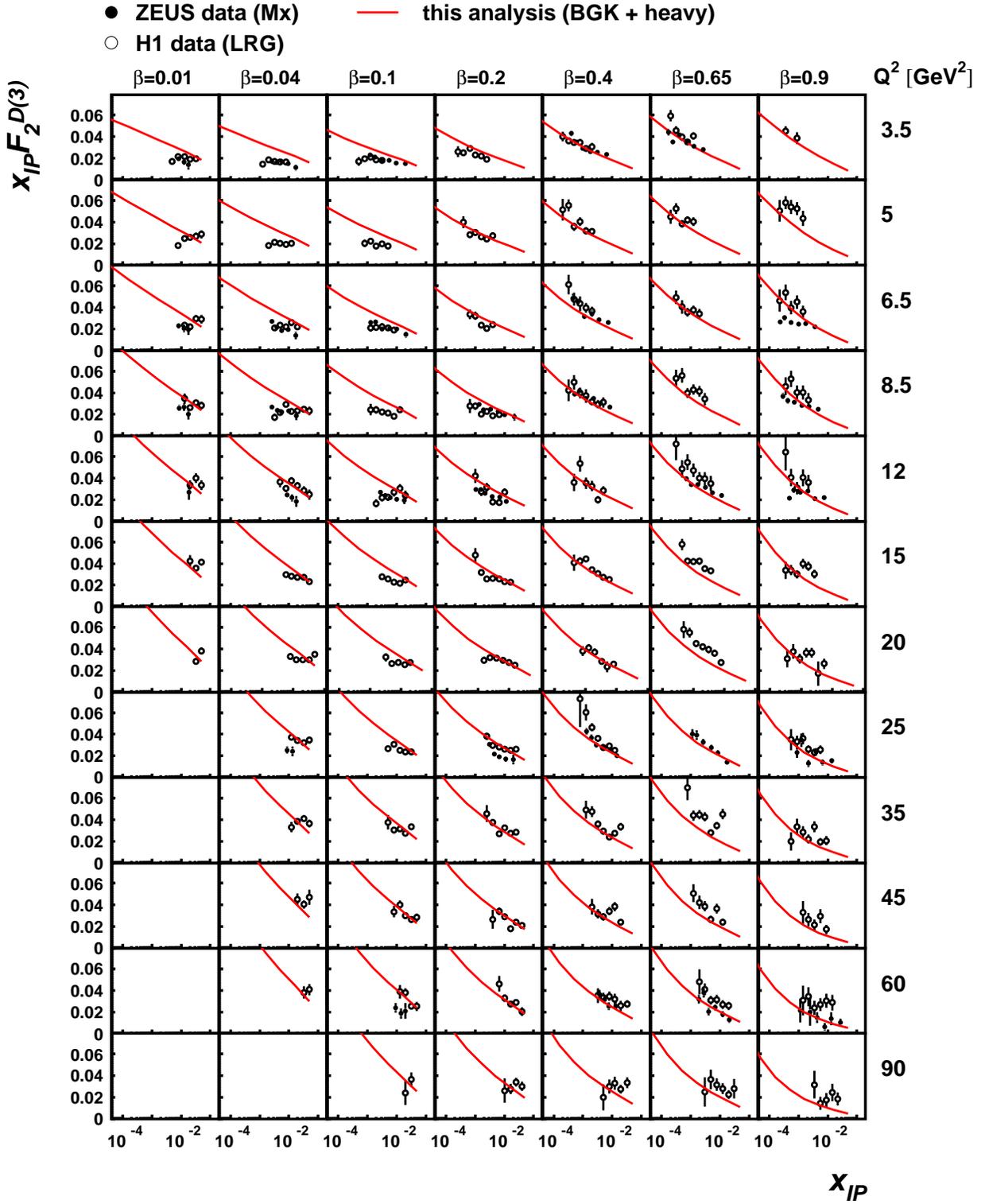,height=8.1in}
\caption{The diffractive structure function $x_P F^{D(3)}(x_P, \beta, Q^2)$. The predictions of our analysis with heavy quarks are compared with the recent H1 and ZEUS data.}
\label{fig:f2diff}
\end{center}
\end{figure}

\vskip 1cm
\centerline{{\bf Acknowledgments}}
We thank the Theory Group members at Ecole Polytechnique where this analysis was completed for their warm hospitality and Laurent Schoeffel for helping us  to prepare Fig.~\ref{fig:f2diff}.
This research has been partly supported by the MEiN research grants: No.~1~P03B~028~28 (2005-08), No.~N202 048 31/2647 (2006-08) and by the French--Polish scientific agreement Polonium.


\end{document}